# Anomalous linear magnetoresistance in high quality crystalline lead thin films

Yi Liu[1,2], Yue Tang[1,2], Ziqiao Wang[1,2], Chaofei Liu[1,2], Cheng Chen[1,2], Jian Wang[1,2,3,4,*]

[1]*International Center for Quantum Materials, School of Physics, Peking University, Beijing 100871, China.*
[2]*Collaborative Innovation Center of Quantum Matter, Beijing 100871, China.*
[3]*CAS Center for Excellence in Topological Quantum Computation, University of Chinese Academy of Sciences, Beijing 1001090, China.*
[4]*Beijing Academy of Quantum Information Sciences, Beijing 100193, China.*

Intriguing novel phenomena in lead films inspire new understanding of quantum physics, such as quantum size effect and quantum phase transitions etc. The improvement of the sample quality makes it even more promising to explore the intrinsic properties in two-dimensional system. In this paper, we show that the crystalline interfacial striped incommensurate layer can increase the quality of the lead films and significantly enhance the magnitude of magnetoresistance. By performing systematic transport measurement, a predominant anomalous linear magnetoresistance is revealed, and the widely used Parish-Littlewood model and Abrikosov's explanation fail to describe the observation. Instead, we propose a new model of linear magnetoresistance based on linear band structure, which shows a good agreement with the experimental results. Our studies reveal a novel origin of linear magnetoresistance which may also be helpful to understand the linear magnetoresistance in other materials with linear dispersion of electronic structure.

Crystalline lead films have been investigated over the past 30 years and gaining growing interest since intriguing phenomena (e.g. quantum size effect [1], superconductivity in the two-dimensional (2D) limit [2-4]) have been continuously discovered and significantly deepen our understanding towards low-dimensional physics [5-10]. One problem influencing the quality of crystalline lead films is the lattice mismatch between films and the substrate. This mismatch, which results in an amorphous interface (named wetting layer), makes it difficult to prepare atomically uniform ultrathin films [11], leading to relatively low mobility and small magnetoresistance (MR) of films. One solution to this mismatch is to grow a crystalline reconstruction phase as the interface between Pb films and Si substrate, such as striped incommensurate (SIC) phase, $\sqrt{7} \times \sqrt{3}$ phase etc. Consequently, several novel phenomena have been observed in ultrathin lead films including interface induced Ising superconductivity [12], anomalous quantum Griffith singularity [13] and so on. These findings well manifest the abundant physical mechanism concealing on the ultrathin crystalline lead films and inspire continuous enthusiasm of the investigations on low-dimensional crystalline systems.

MR measurement is a typical and useful method to reveal the physical properties including density of states (DOS), mobility and band structures. One striking phenomenon is the linear magnetoresistance (LMR), which could be observed in some special situations and may survive up to very high magnetic field [14-16] (e.g. LMR in $Cd_3As_2$ crystal can survive up to 60 T). In theory, Parish-Littlewood (PL) model [17] and Abrikosov's quantum explanation [18,19] are widely used to describe the LMR. The classical PL model contributes the LMR in disordered system to the large inhomogeneity, which distorts the current flow and introduces transverse Hall resistance to longitudinal conducting compositions. Another mechanism of LMR was found by Abrikosov, which ascribes LMR to the intrinsic quantum states of systems. If the system reaches the "extreme quantum limit", which means that the electrons all stay at the first Landau level, the LMR would appear naturally according to Abrikosov's quantum explanation [18]. However, not all the cases could be well explained by these two models such as the LMR in iron-based superconductor [20], Dirac semimetal [21] and ferromagnetic semiconductor [22]. Several novel mechanisms have been proposed including spin fluctuation and d band shift [22]. In order to reach a deeper and wider understanding of LMR in various systems, further experiments and explanations are highly desired.

In this paper, we report the transport measurements on 10-mololayer (ML) and 20- ML crystalline lead films grown on SIC phase on Si(111) substrate via molecular beam epitaxy (MBE). Compared to the lead film grown on amorphous wetting layers, the 20-ML Pb film grown on SIC layer exhibit a significantly larger mobility. As a result, a large enhancement of MR is observed in this system. To be specific, the MR, defined as $MR = [R(H) - R(0)]/R(0)$, reaches 3% and 11% under 15T at 10K in 10-ML and 20-ML Pb films with SIC phase interface. Pb thin films follow classical quadratic field dependence in low magnetic field regime, however, a predominant LMR appears at 5T and exists up to 15T. The widely used PL model and Abrikosov's explanation fail to explain our observations. We ascribe the detected LMR in 2D films to the changes of DOS for linear bands at the Fermi surface.

Figure 1(a) presents the morphology and atomically resolved image of 20-ML lead film measured by scanning tunneling microscopy (STM), indicating the high quality of samples grown on SIC phase. The transport properties at zero magnetic field of 20-ML Pb film are shown in Fig. 1(b). A relatively large residual resistance ratio (RRR) of 13 (defined as the ratio of resistance at 300K and 8K, RRR=R(300K)/R(8K)) is observed, which is twice as much as that of the samples grown on amorphous wetting layers [Fig. S2][23], showing a great improvement of sample quality by using SIC phase interface. The schematic of standard four-electrode transport measurement is presented in the inset of Fig. 1(b). Figure 1(c) and 1(d) summarize the perpendicular field dependence of longitudinal resistance for both 10-ML and 20-ML lead films at various temperatures. The MR monotonically decreases with increasing temperature, which can be ascribed to the decreasing mobility at higher temperatures [Fig. S3][23]. Under low magnetic field, MR increases quadratically, and strikingly, as the field exceeds 5T, MR gradually turns to the linear behavior. The change from parabolic MR to LMR could be shown more clearly in the field dependent dMR/dB (Fig. 1(e) and 1(f)). The intercept of dMR/dB vs B curve represents the linear terms (MR $\propto B$) while the gradient refers to the quadratic component (MR $\propto B^2$) (see Eq.(1)). The intercepts of the curves in the high field regime remains finite below 30K indicating the survival of LMR. The LMR at relatively low temperatures (large intercept and small slope) gradually evolves to quadratic behavior (small intercept and large slope) at higher temperatures, revealing the appearance of parabolic term and the disappearance of LMR with increasing temperatures. In Fig. 1(e) and 1(f), the curves could be separated to three different parts. The first part is a straight line passing the origin at low fields (quadratic term), the second is a smooth transition region and the last is a straight line with a reduced gradient (linear and quadratic terms). The MR in the first and the last parts could be expressed as [24]:

$$\frac{dMR}{dB} = \begin{cases} 2A_0\mu^2 B, & B < B_c \\ A_1\mu + 2A_2\mu^2 B, & B \geq B_c \end{cases} \quad (1)$$

$$MR = \begin{cases} A_0\mu^2 B^2, & B < B_c \\ A_1\mu B + A_2\mu^2 B^2, & B \geq B_c \end{cases} \quad (2)$$

where $A_0$, $A_1$ and $A_2$ are constant parameters, $\mu$ is the mobility and the crossover field $B_c$ corresponds to the crossing point of two black solid lines. Eq. (2) includes the parabolic term of MR with the coefficient $A_0$ under low field and a linear component plus a quadratic term with the coefficients $A_1$ and $A_2$ at high fields, respectively. As shown in Fig. 2, the values of $A_0$ and $A_1$ generally decreases and $A_2$ increases with warming temperature. The increasing $A_2$ is quite distinct from the previous study showing a deceasing coefficient of the quadratic term when the temperature is increasing [24]. In the Pb films, $A_2$ (the quadratic term in high field region) increases with increasing temperature and finally be almost equal to $A_0$ (the quadratic term in low field region) indicating the disappearance of LMR.

Figure 3 shows the temperature dependence of crossover field extracted from Fig. 1(e) and 1(f) for both 20-ML and 10-ML Pb films (see Fig. S11 for the extraction of $B_c$ above 30K for the 20-ML Pb film [23]). The crossover field monotonically increases with increasing temperature and shows a tendency of saturation around 50 K. As shown in Fig. 3, the crossover field could be well fitted by a thermal activation equation (the red lines in Fig. 3) which reads $B_c = A \cdot e^{-\frac{\Delta}{k_B T}} + C$, where $\Delta$ is the activation energy, $A$, and $C$ are coefficients and $k_B$ is Boltzmann constant. The relatively large DOS in the Pb films makes the quantum limit much larger than 15 T, thus the Abrikosov's quantum scenario is not consistent with our observation [23]. The classical PL model predicts that LMR more likely occurs in more disordered system, however, 20-ML Pb film with relatively high mobility exhibits a clearer LMR compared to the 10-ML Pb film with low mobility. Besides, the behaviors of $1/\mu$ vs $B_c$ curves are found to contradict the expectation of the PL model [23]. Moreover, the theories based on the orbital effects and the scattering of the cyclotron electrons [25-27] can also be excluded due to the observation of LMR under parallel field (detailed discussions are in Ref. [23]).

To understand the observed LMR, we propose a new phenomenological model based on the linear bandstructure. In our model, we assume a small region of Dirac-type band (linear dispersion) around Fermi surface and other bands are the normal parabolic bands resulting in quadratic behavior of MR. Figure 4(a) presents the DOS versus energy of the aforementioned band structure. In 2D systems, the DOS $N(E)$ of parabolic band is independent of energy (the vertical solid line in Fig. 4(a)) while the $N(E)$ of Dirac-type band linearly changes with increasing energy values (the solid line with finite slope in Fig. 4(a)). The up and down arrows in x axis represent the two opposite directions of spins and the y axis separates the band structure with different spin directions. The blue area represents the occupied electron states and the band structure takes a hole-type shape since hole is predominant in lead films. When external magnetic field is increasing, Zeeman splitting would lead to the opposite shift of the band structures and the electrons in the elevated bands would naturally tend to lower their energy and thus jump to the states with opposite spin directions (illustrated by the curved arrow in Fig. 4(a)). In Fig. 4(a), the spin splitting is linear to the external magnetic field ($\delta h = g\mu_B B$, where $g$ is Lande factor and $\mu_B$ is Born magneton), and leads to the slight increase of the Fermi level (black dashed line). Thus, the DOS in the Fermi level also decreases (marked as $\delta N(E_F)$) and consequently affects the value of the conductance. The change of DOS is proportional to the magnetic field $\delta N(E_F) = -\gamma g\mu_B B$ where $\gamma$ is a constant determined by the shape of band structures. In 2D systems the conductance can be expressed as $\sigma = \frac{1}{2} j_F^2 \tau(E_F) N(E_F)$ [23]. Thus, the MR can be deduced as

$$\frac{\delta\rho(B)}{\rho_0} = -\frac{\delta\sigma(B)}{\sigma_0} = -\frac{\delta(\tau(E_F)N(E_F))}{\tau_0 N_0} = -\frac{\tau(E_F)\delta N(E_F)}{\tau_0 N_0} + \tau(E_F)\delta\left(\frac{1}{\tau}\right)\frac{N(E_F)}{N_0} \qquad (3)$$

where $j_F$ is a constant representing the Fermi-surface current density and $\tau$ is the relaxation rate. Pb film is a typical s-wave superconductor where the electron-phonon interaction is dominant. Therefore, the relaxation rate of Pb thin films can be understood by deformation potential theory [28,29] and thus can be expressed as $\frac{1}{\tau} = \frac{1}{\tau_0} + \frac{1}{\tau_{de}} = \frac{1}{\tau_0} + K(T) \cdot N(E)$ [29-31] where $\tau_0$ is the term independent with magnetic field and $K(T)$ is a temperature-dependent function. After applying this function into Eq. (3), the LMR can be deduced clearly:

$$MR = \frac{\delta\rho(B)}{\rho_0} = -\frac{\delta N(E_F)}{N_0}\left(\frac{1}{1+KN_0\tau_0}\right)^2 \propto \frac{B}{N_0 \cdot (1+KN_0\tau_0)^2} \qquad (4)$$

Interestingly, the crossover from the low field quadratic MR to high field LMR can also be explained in the framework of our phenomenological model. In Fig. 4(c), the Fermi surface is initially in the quadratic band area. When the external field slightly shifts the DOS of the bands (corresponding to the small magnetic field regime) and the electrons have not reached the linear band regime, the DOS at the Fermi surface remains a constant and cannot give rise to the LMR [Fig. 4(d)]. Only when the field is large enough to make the electron occupy the states in the linear band, the DOS begin to decrease and lead to the LMR [Fig. 4(e)]. The field required to reach the border of the linear bands is the crossover field $B_c$. Furthermore, the observed thermally activated behavior of $B_c$ [Fig. 3] can be understood as follows. At finite temperatures, the thermal activation would bring some of the lower band's electrons to the upper band [Fig. 4(b)]. The number of such electrons shows an activation function relationship with the temperature $\delta n \propto \exp(-\frac{\Delta}{k_B T})$ where $\Delta$ describes the gap between lower and upper bands. Since the number of electrons decreases at lower bands, a higher magnetic field is required to make the lower band's electron reach the border of linear bands. Therefore, the crossover field also shows a thermal activation behavior. Indeed, the data in Fig. 3 can be well fitted by the thermal activation function $B_c = A \cdot e^{-\frac{\Delta}{k_B T}} + C$ with the activation energy of 1.4 meV and 1.0 meV for 20-ML and 10-ML Pb films, respectively. The small activation energy could exist in the lead thin films due to the intricate band structures [12].

The expectations of our model are well consistent with the experimental data. The main assumption of this model is the existence of a linear band in a small region near the Fermi level, which has been indicated by previous works, including bandstructure calculations and angle-resolved photoemission spectroscopy (ARPES) measurements. The APRES data of 21-ML Pb films along the ΓM direction show linear bands away from the Γ point and parabolic bands around the Γ point, which is confirmed by the bandstruture calculations [32]. Further investigations on 23-ML and 24-ML Pb films along the ΓK direction reveal similar bandstructure with linear energy dispersion near the Fermi surface [33]. The emergence of linear bands near the Fermi level is also widely observed in thinner films, such as 6-ML to 10-ML Pb films [34-37]. Therefore, the bandstruture of Pb thin films exhibits both parabolic and linear bands, which is well consistent with our model. Moreover, the prevailing classical PL model explains the LMR in disordered systems [17,38,39], while the Abrikosov's quantum explanation requires the quantum limit [18], which is very hard to reach for most materials. Different from them, our work proposes a new model of LMR, which is not only a new understanding towards LMR, but also inspires further investigations on the linear MR in 2D systems with linear energy dispersion (e.g. the topological materials and high-temperature superconductors FeSe and FeTe$_{1-x}$Se$_x$ etc), especially for those with less disorder and relatively high carrier density.

In summary, with the improvement of sample quality, we performed detailed magnetic transport measurements on both 10-ML and 20-ML crystalline Pb films. Surprisingly, a pronounced LMR is detected from 8 K to 30 K in both films and cannot be interpreted by the widely accepted theories, i.e. the classical PL model and the Abrikosov's quantum explanation. With an attempt to understand the concealing mechanism of the observed LMR, we propose a phenomenological model based on the band structure composed of linear and quadratic bands, which suggests that the observed LMR is originated from the changes of DOS at the Fermi surface. The model can well describe the crossover from quadratic MR to LMR and explain the temperature dependence of $B_c$. With systematic studies of MR in Pb thin films, our work reveals a new mechanism of LMR, which could be applied to various 2D materials with linear energy dispersion.

We thank Prof. Haiwen Liu for the fruitful discussion and Prof. Ying Xing for the help in transport measurement. This work was financially supported by the National Key Research and Development Program of China (Grant No. 2018YFA0305604, No. 2017YFA0303302), the National Natural Science Foundation of China (Grant No. 11888101, No.11774008), the Strategic Priority Research Program of Chinese Academy of Sciences (Grant No. XDB28000000), Beijing Natural Science Foundation (Z180010) and China Postdoctoral Science Foundation (Grant No. 2019M650290 and No. 2020T130021).

Y. Liu and Y. Tang contributed equally to this work.
*Corresponding author. Jian Wang (jianwangphysics@pku.edu.cn)

[1] Y. Guo, Y.-F. Zhang, X.-Y. Bao, T.-Z. Han, Z. Tang, L.-X. Zhang, W.-G. Zhu, E. G. Wang, Q. Niu, Z. Q. Qiu, J.-F. Jia, Z.-X. Zhao, and Q.-K. Xue, Science **306**, 1915 (2004).

[2] T. Zhang, P. Cheng, W.-J. Li, Y.-J. Sun, G. Wang, X.-G. Zhu, K. He, L. Wang, X. Ma, X. Chen, Y. Wang, Y. Liu, H.-Q. Lin, J.-F. Jia, and Q.-K. Xue, Nat. Phys. **6**, 104 (2010).

[3] S. Qin, J. Kim, Q. Niu, and C.-K. Shih, Science **324**, 1314 (2009).

[4] M. Yamada, T. Hirahara, and S. Hasegawa, Phys. Rev. Lett. **110**, 237001 (2013).

[5] H. H. Weitering, D. R. Heslinga, and T. Hibma, Phys. Rev. B **45**, 5991 (1992).

[6] M. M. Özer, J. R. Thompson, and H. H. Weitering, Nat. Phys. **2**, 173 (2006).

[7] D. Eom, S. Qin, M. Y. Chou, and C. K. Shih, Phys. Rev. Lett. **96**, 027005 (2006).

[8] C. Brun, T. Cren, V. Cherkez, F. Debontridder, S. Pons, D. Fokin, M. C. Tringides, S. Bozhko, L. B. Ioffe, B. L. Altshuler, and D. Roditchev, Nat. Phys. **10**, 444 (2014).

[9] Y. Saito, T. Nojima, and Y. Iwasa, Nat. Rev. Mater. **2**, 16094 (2016).

[10] C. Brun, I. P. Hong, F. Patthey, I. Y. Sklyadneva, R. Heid, P. M. Echenique, K. P. Bohnen, E. V. Chulkov, and W.-D. Schneider, Phys. Rev. Lett. **102**, 207002 (2009).

[11] Y.-F. Zhang, J.-F. Jia, Z. Tang, T.-Z. Han, X.-C. Ma, and Q.-K. Xue, Surf. Sci. **596**, L331 (2005).

[12] Y. Liu, Z. Wang, X. Zhang, C. Liu, Y. Liu, Z. Zhou, J. Wang, Q. Wang, Y. Liu, C. Xi, M. Tian, H. Liu, J. Feng, X. C. Xie, and J. Wang, Phys. Rev. X **8**, 021002 (2018).

[13] Y. Liu, Z. Wang, P. Shan, Y. Tang, C. Liu, C. Chen, Y. Xing, Q. Wang, H. Liu, X. Lin, X. C. Xie, and J. Wang, Nat. Commun. **10**, 3633 (2019).

[14] Y. Zhao, H. Liu, C. Zhang, H. Wang, J. Wang, Z. Lin, Y. Xing, H. Lu, J. Liu, Y. Wang, S. M. Brombosz, Z. Xiao, S. Jia, X. C. Xie, and J. Wang, Phys. Rev. X **5**, 031037 (2015).

[15] T. Khouri, U. Zeitler, C. Reichl, W. Wegscheider, N. E. Hussey, S. Wiedmann, and J. C. Maan, Phys. Rev. Lett. **117**, 256601 (2016).

[16] X. Zhang, T. Luo, X. Hu, J. Guo, G. Lin, Y. Li, Y. Liu, X. Li, J. Ge, Y. Xing, Z. Zhu, P. Gao, L. Sun, and J. Wang, Chinese Phys. Lett. **36**, 057402 (2019).

[17] M. M. Parish and P. B. Littlewood, Nature **426**, 162 (2003).

[18] A. A. Abrikosov, Phys. Rev. B **58**, 2788 (1998).

[19] A. A. Abrikosov, Europhys. Lett. **49**, 789 (2000).

[20] Q. Wang, W. Zhang, W. Chen, Y. Xing, Y. Sun, Z. Wang, J.-W. Mei, Z. Wang, L. Wang, X.-C. Ma, F. Liu, Q.-K. Xue, and J. Wang, 2D Mater. **4**, 034004 (2017).

[21] M. Novak, S. Sasaki, K. Segawa, and Y. Ando, Phys. Rev. B **91**, 041203 (2015).

[22] K. J. Kormondy, L. Gao, X. Li, S. Lu, A. B. Posadas, S. Shen, M. Tsoi, M. R. McCartney, D. J. Smith, J. Zhou, L. L. Lev, M.-A. Husanu, V. N. Strocov, and A. A. Demkov, Sci. Rep. **8**, 7721 (2018).

[23] See Supplemental Material for details.

[24] Y. Sun, S. Pyon, and T. Tamegai, Phys. Rev. B **93**, 104502 (2016).

[25] A. A. Sinchenko, P. D. Grigoriev, P. Lejay, and P. Monceau, Phys. Rev. B **96**, 245129 (2017).

[26] Y. Feng, Y. Wang, D. M. Silevitch, J. Q. Yan, R. Kobayashi, M. Hedo, T. Nakama, Y. Ōnuki, A. V. Suslov, B. Mihaila, P. B. Littlewood, and T. F. Rosenbaum, PNAS **116**, 11201 (2019).

[27] B. Wu, V. Barrena, F. Mompeán, M. García-Hernández, H. Suderow, and I. Guillamón, Phys. Rev. B **101**, 205123 (2020).

[28] J. Bardeen and W. Shockley, Phys. Rev. **80**, 72 (1950).

[29] J. M. Ziman, *Principles of the Theory of Solids* (Cambridge. Univ. Press, 1972).

[30] M. T. Dylla, S. D. Kang, and G. J. Snyder, Angew. Chem. Int. Ed. **58**, 5503 (2019).

[31] H. Goldsmid, *Thermoelectric refrigeration* (Springer, 2013).

[32] S. He, Z. Zeng, M. Arita, M. Sawada, K. Shimada, S. Qiao, G. Li, W.-X. Li, Y.-F. Zhang, Y. Zhang, X. Ma, J. Jia, Q.-K. Xue, H. Namatame, and M. Taniguchi, New J. Phys. **12**, 113034 (2010).

[33] Y. J. Sun, S. Souma, W. J. Li, T. Sato, X. G. Zhu, G. Wang, X. Chen, X. C. Ma, Q. K. Xue, J. F. Jia, T. Takahashi, and T. Sakurai, Nano Res. **3**, 800 (2010).

[34] J. H. Dil, J. W. Kim, T. Kampen, K. Horn, and A. R. H. F. Ettema, Phys. Rev. B **73**, 161308 (2006).

[35] N. Miyata, K. Horikoshi, T. Hirahara, S. Hasegawa, C. M. Wei, and I. Matsuda, Phys. Rev. B **78**, 245405 (2008).

[36] J. H. Dil, F. Meier, J. Lobo-Checa, L. Patthey, G. Bihlmayer, and J. Osterwalder, Phys. Rev. Lett. **101**, 266802 (2008).

[37] M. Müller, N. Néel, S. Crampin, and J. Kröger, Phys. Rev. B **96**, 205426 (2017).

[38] R. Xu, A. Husmann, T. F. Rosenbaum, M. L. Saboungi, J. E. Enderby, and P. B. Littlewood, Nature **390**, 57 (1997).

[39] Z. H. Wang, L. Yang, X. J. Li, X. T. Zhao, H. L. Wang, Z. D. Zhang, and X. P. A. Gao, Nano Lett. **14**, 6510 (2014).

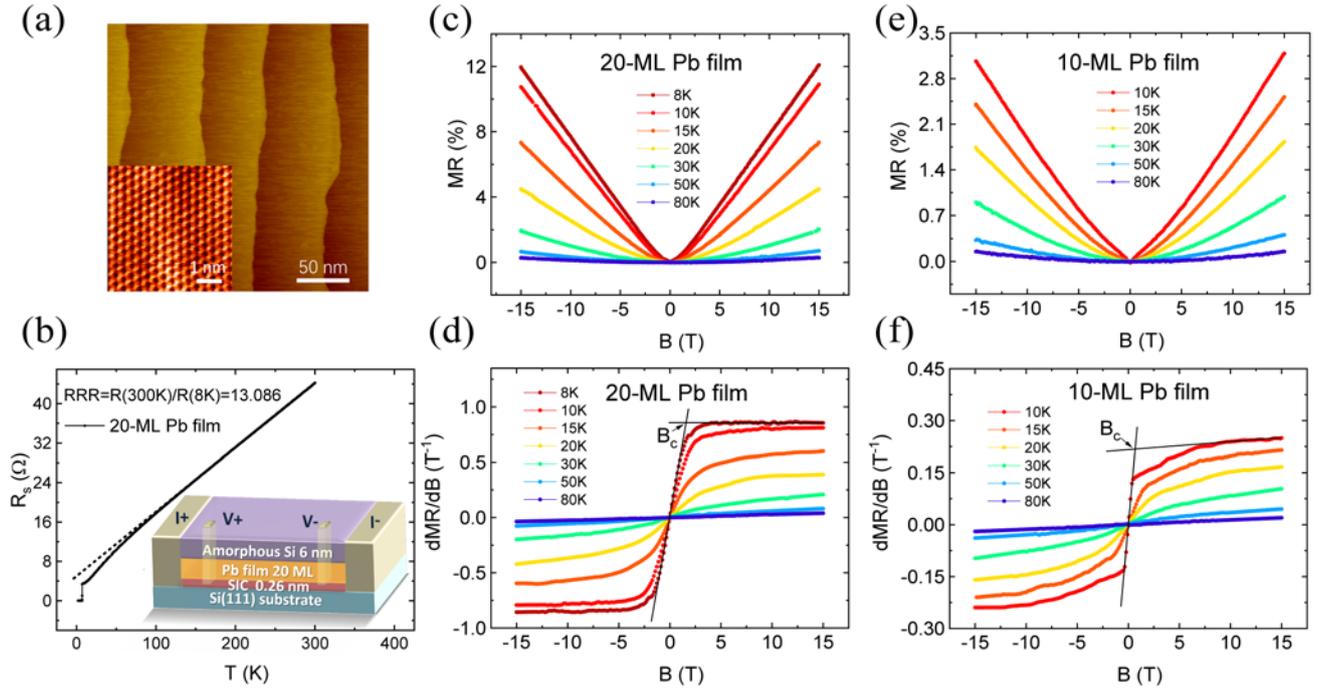

FIG. 1. Morphology and transport properties of Pb thin films. (a) A typical STM image of 20-ML Pb film (250 nm × 250 nm). The inset: atomically resolved STM image of 20-ML Pb film (5 nm × 5 nm). (b) Temperature dependence of sheet resistance on 20-ML Pb film at zero magnetic field, showing RRR of 13.086 defined as the ratio of resistance at 300K and 8K (RRR=R(300K)/R(8K)). The inset is a schematic diagram for standard four-electrode transport measurements. Perpendicular magnetic field dependence of sheet resistance for (c) 20-ML and (d) 10-ML Pb films on SIC phase interface. The corresponding derivative MR (dMR/dB) is shown in (e) and (f), respectively.

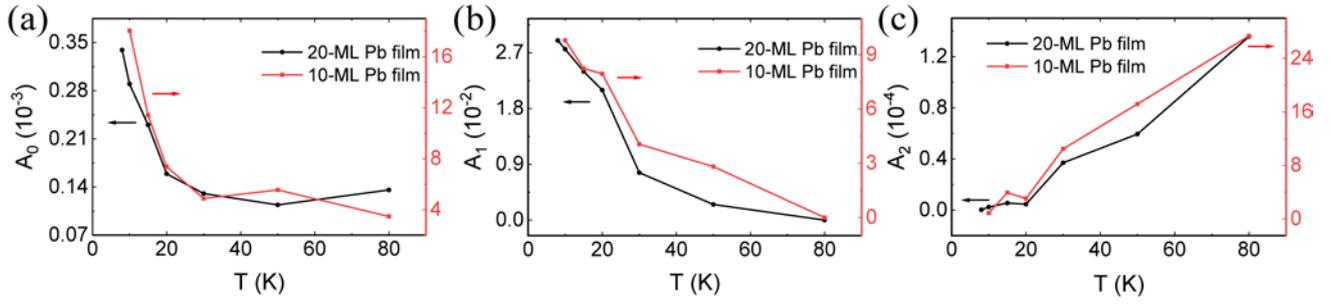

FIG. 2. The temperature dependence of (a) low-field quadratic term coefficient $A_0$, (b) high-field linear term coefficient $A_1$ and (c) high-field quadratic term coefficient $A_2$ for 10-ML and 20-ML Pb films. All the coefficients are obtained from the fitting results of Eq. (2).

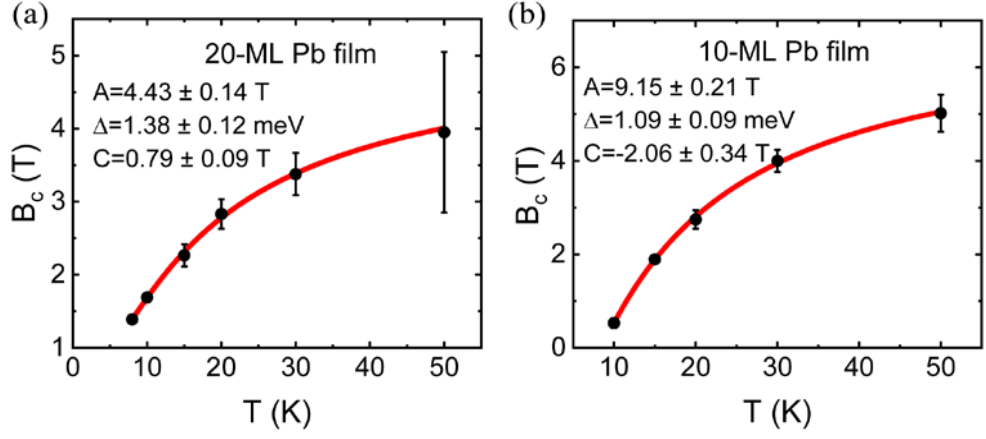

FIG. 3. The temperature dependence of crossover field $B_c$ for both (a) 20-ML and (b) 10-ML films. The red lines are theoretical fitting curves ($B_c = A \cdot e^{-\frac{\Delta}{k_B T}} + C$). The fitting parameters $A, C$ and $\Delta$ are summarized in this figure.

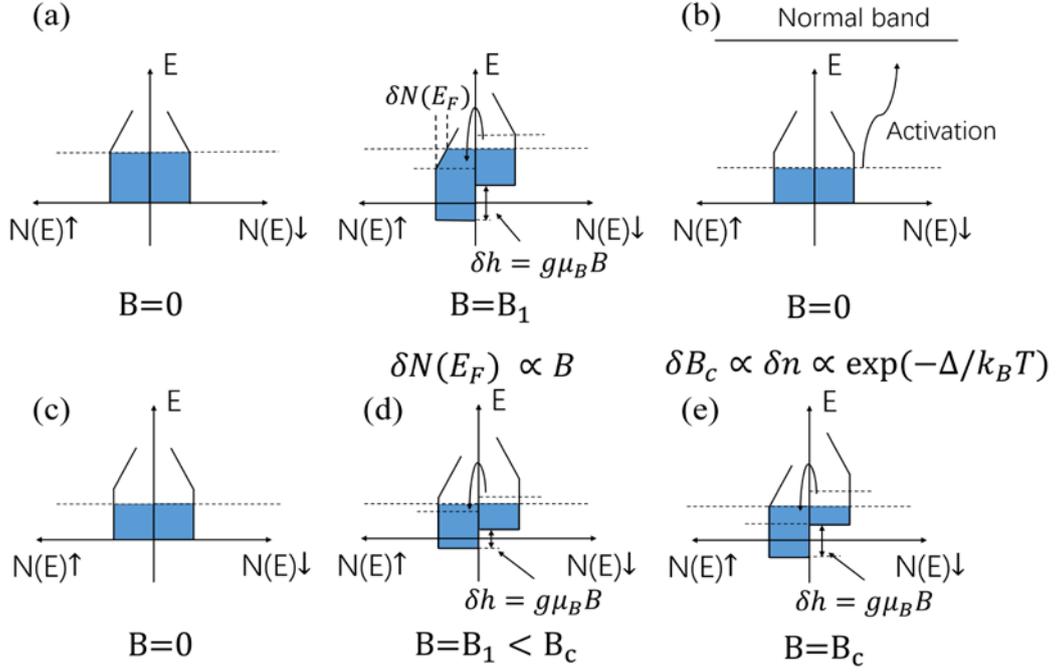

FIG. 4. The illustrations for the origin of LMR. (a) The DOS dependence of energy in the band structure consisting of linear (the oblique lines) and quadratic bands (the vertical lines) in 2D system. The blue color represents the states occupied by electrons and the horizontal dashed line represents the Fermi surface while the vertical dashed lines represent the change of DOS, $\delta N(E_F)$, at the Fermi level. The change of DOS at the Fermi surface leads to LMR. (b) the illustration of thermally activated behavior of crossover field. (c-e) The illustration of appearance of LMR. Only when the magnetic field makes the Fermi surface reach the border of linear bands, the DOS would start a relatively large change leading to the appearance of LMR.

Supplemental Material for:

# Anomalous linear magnetoresistance in high quality crystalline lead thin films

Yi Liu[1,2], Yue Tang[1,2], Ziqiao Wang[1,2], Chaofei Liu[1,2], Cheng Chen[1,2], Jian Wang[1,2,3,4,*]

[1]*International Center for Quantum Materials, School of Physics, Peking University, Beijing 100871, China.*
[2]*Collaborative Innovation Center of Quantum Matter, Beijing 100871, China.*
[3]*CAS Center for Excellence in Topological Quantum Computation, University of Chinese Academy of Sciences, Beijing 1001090, China.*
[4]*Beijing Academy of Quantum Information Sciences, Beijing 100193, China.*

**Contents**

I. Methods
II. The discussion on Parish-Littlewood (PL) model, Abrikosov's explanation and other theoretical models
III. The theoretical explanation of linear magnetoresistance (LMR) in Pb thin films
IV. The estimation of temperature-dependent mobility
V. The discussion on 20-ML Pb films grown on wetting layer and striped incommensurate (SIC) phase interface
VI. Figures

## I. Methods

**Sample growth.** The 10-monolayer (ML) and 20-ML crystalline Pb (111) thin films were grown by molecular beam epitaxy in an ultrahigh vacuum chamber (Omicron) with a base pressure lower than $1\times10^{-10}$ mbar. The Si(111) substrate was cyclically flashed at T ~ 1400 K to prepare the Si(111)- 7×7 reconstruction phase. After depositing 1.5 ML Pb from a Knudsen cell at room temperature, the striped incommensurate Pb phase was prepared by annealing the sample at T ~ 573 K for 30 sec. The crystalline Pb films were grown by depositing pure Pb atoms on the SIC phase or amorphous wetting layer at 150 K with a growing rate of ~0.2 ML/min. Film growth process was monitored by reflection high-energy diffraction (RHEED) and the sample quality was characterized by scanning tunneling microscopy (STM).

**Transport measurement.** The Pb thin films were protected by 6-nm thick amorphous Si capping layer for *ex situ* transport measurements. The resistance and magnetoresistance of the films were measured using the standard four-probe method in a commercial Physical Property Measurement System (Quantum Design, PPMS-16) for temperatures down to 2 K and perpendicular magnetic fields up to 15 T. The contacts between electrodes and films were made by indium electrodes pressed directly on the films.

## II. The discussion on PL model, Abrikosov's explanation and other theoretical models

In order to understand the origin of LMR in Pb films, we firstly consider two widely accepted theories, Abrikosov's explanation [1,2] and PL model [3]. Abrikosov's explanation predicts the existence of LMR when the system stays in the "extreme quantum limit" that means all electrons should stay at the first Landau level. To satisfy this condition, since the degeneracy of the first Landau level is finite, the number of electrons in the system shall be limited and the temperature should be also constrained due to the broadening of Landau level. Therefore, there are two inequalities as conditions for LMR which can be expressed as [2]:

$$n_0 \ll \left(\frac{eB}{\hbar}\right)^{\frac{3}{2}}, \qquad T \ll \frac{eB\hbar}{m^* k_B}$$

where $n_0$ is density of carrier; $m^*$ is the effective mass. For systems possessing linear band structures such as graphene, topological insulator, Dirac and Wely semimetals, the LMR is usually explained by Abrikosov's explanation because the effective mass of linear bands is zero that naturally satisfies the second inequality. However, the observed LMR in Pb films cannot be explained by this theory because of their large carrier density. The carrier density in our systems is larger than $10^{23} cm^{-3}$ (Fig. S3(a) and Fig. S3(b)) which greatly exceeds the value of the right side of the first inequality that is around $10^{18} cm^{-3}$ under 10T. In this case, the large number of carriers have already filled up the first Landau level and therefore rule out the most expected explanation improved by Abrikosov.

Different from the Abrikosov's quantum explanation, PL model contributes LMR to the large inhomogeneity of the system. The large inhomogeneity, according to PL model's interpretation, would distort the current flow and hence introduce the transverse Hall component into the longitudinal resistance which would finally give rise to the linear response to the external magnetic field. This theory is usually applied to the systems with large granularity or large mobility fluctuation [4-6], which seems not suitable to the crystalline Pb films. For the systems with low mobility fluctuation $\Delta\mu < \mu_{average}$, PL model predicts that the crossover field is equal to the inverse of the average of mobility, $B_c = \langle\mu\rangle^{-1}$ [4,7]. However, the mobility of our films are smaller than $30 \ cm^2/V \cdot s$ (Fig. S3), which means that a magnetic field more than 300T is required to satisfy the PL model's expectation. This extremely large field obviously contradicts the fact that LMR could just appear over around 5 T in Pb thin films. The curve of $1/\mu$ versus $B_c$ also does not show a linear relationship (Fig. S3(c) and Fig, S3(d)). Moreover, PL model predicts that under high magnetic field it should be satisfied that $dMR_{linear}/dB \propto \mu$ [4,8] (i.e. $A_1$ is a constant independent of temperature). However, in Fig. 2(a) of main text, it is shown that the $A_1$ is monotonically decreasing with increasing temperature. In addition, the LMR that could be explained by PL model could usually survive up to room temperature [5,6], which is not consistent with our observation that LMR disappears around 30K. Therefore, PL model cannot explain our observation of LMR in Pb thin films.

Moreover, the orbital effects may play an important role under perpendicular field and should be discussed. According to previous theories[9,10], the orbital effects are pronounced for a 2D system when $\mu \cdot B > 1$, where $\mu$ represents the mobility. The mobilities of 20-ML and 10-ML Pb films are smaller than $30 \ cm^2/V \cdot s$ (Fig. S3), which requires the magnetic field larger than 333 T. However, the largest magnetic field in our measurement is 15 T, which indicates that $\mu \cdot B \ll 1$. Therefore, the orbital effects are unlikely to be dominant in our crystalline Pb thin films. Furthermore, Fig. S7 shows the transport measurement of 20-ML Pb film on SIC phase under parallel magnetic field. A dominant LMR surviving up to 15T is observed, which cannot be ascribed to the orbit effects. The magnitude of LMR under a parallel field (2.5% at 8K under 15T) is smaller than that under a perpendicular field (12% at 8K under 15T). There are several parameters determining the magnitude of the LMR under different field directions. For example, in our model, if the Laude g factor in a perpendicular field is larger than that in a parallel field, the LMR should be more pronounced under perpendicular field.

Another possible origin of LMR is the "hot spots" theory, which involves the charge density wave (CDW) [11-13]. According to this theory, when there are no open orbits in the system, the scattering of cyclotron electrons on CDW fluctuations is strongest near the "hot spots" on the Fermi surface, which results in a different scattering time and finally gives rise to the LMR. However, there are lacking experimental reports and theoretical predictions about the CDW transition as well as open orbits in the Pb thin films even though the physical properties of Pb films have been investigated for around 80 years. Furthermore, the LMR surviving up to 15T is also observed under parallel magnetic field. The cyclotron electrons do not exist under parallel field for Pb thin films where the electrons are in quantum well states. Therefore, this "hot spots" theory related to the scattering of cyclotron electrons cannot explain our experimental results.

Last but not least, we present the Kohler plot of 20-ML and 10-ML Pb films under perpendicular magnetic field (Fig. S6). Clearly, the curves at different temperatures cannot be collapsed to a universe curve, which means that below 80K Kohler rule cannot be applied to our Pb films under perpendicular field. Semiclassical transport theory predicts Kohler's rule to hold if there is a single type of charge carrier and the scattering/relaxation time τ keeps the same at the whole Fermi surface under different magnetic fields. The inapplicability of the Kohler rule is consistent with the model we proposed since our model assumes a scattering time dependent on the external magnetic field.

### III. The theoretical explanation of LMR in Pb thin films

Based on the traditional analysis of the conductance [14], the basic formula of the current operator reads:

$$\vec{J} = 2\int e\vec{v_k}g_k d^3k, \qquad g_k = \left(-\frac{\partial f^0}{\partial \epsilon}\right)\tau\vec{v_k}\cdot e\vec{E}$$

where $\vec{v_k}$ is the velocity operator, $f$ is the distribution function of electrons, $\tau$ is the relaxation time and $\vec{E}$ refers to the external electric field.

$$J = \int e^2\tau\vec{v_k}(\vec{v_k}\cdot\vec{E})\left(-\frac{\partial f^0}{\partial \epsilon}\right)N(\epsilon)d\epsilon = \int e^2\tau\vec{v_k}\,\vec{v_k}\cdot\vec{E}\delta(\epsilon-\epsilon_F)N(\epsilon)d\epsilon$$

Here, we consider the case around zero temperature, where the differentiation of distribution function, $-\frac{\partial f^0}{\partial \epsilon}$, could be approximately regarded as the Dirac function, $\delta(\epsilon-\epsilon_F)$. Besides, we deal with crystals with cubic symmetry and for simplicity, we think of the case where the vectors, $\vec{E}$ and $\vec{J}$, are both along the x direction. In this case, then,

$$\left(\vec{v_k}\,\vec{v_k}\cdot\vec{E}\right)_x = v_x^2\,E$$

Since in two-dimensional system $v_x^2$ is 1/2 of total velocity, we obtain:

$$\sigma = \frac{1}{2}j_F^2\tau(\epsilon_F)N(\epsilon_F) \qquad (1)$$

where $j_F = ev_F$ describes the density of the Fermi current.

Based on Eq.(1), the magnetoresistance could also be expressed by the density of state at the Fermi surface

$$\frac{\delta\rho(B)}{\rho_0} = -\frac{\delta\sigma(B)}{\sigma_0} = -\frac{\delta(\tau(\epsilon_F)N(\epsilon_F))}{\tau_0 N_0} = -\frac{\delta N(\epsilon_F)}{N_0}\frac{\tau(\epsilon_F)}{\tau_0} + \frac{\tau^2}{\tau_0}\delta\left(\frac{1}{\tau}\right)\frac{N(\epsilon_F)}{N_0} \qquad (2)$$

The second term in Eq.(2) is related to the relaxation time $\tau$ as which we regard the explanation of deformation potential theory [14,15]. In this theory, the phonon-electron scattering is considered as the predominant effect and its scattering probability is proportional to the density of state $N(\epsilon)$. The inverse of the relaxation time is also proportional to the scattering probability. Therefore, the relaxation time could be expressed as:

$$\frac{1}{\tau_{deformation}} = K\cdot N(\epsilon) \qquad (3)$$

Where $K$ is a constant independent of magnetic field.

Deformation potential theory stresses the phonon-electron scattering, however, in a real system, the other scattering mechanism, such as scattering between impurities and electrons, should also be taken into consideration. We assume these effects as an effective efficient $\tau_0$ and express the total relaxation time as:

$$\frac{1}{\tau} = \frac{1}{\tau_0} + \frac{1}{\tau_{deformation}} = \frac{1}{\tau_0} + K \cdot N(\epsilon) \quad (4)$$

Applying Eq.(4) to Eq.(2), the relationship between magnetoresistance and the density of state around zero temperature could read:

$$\frac{\delta\rho(B)}{\rho_0} = -\frac{\delta N(\epsilon_F)}{N_0}\left(\frac{\tau}{\tau_0}\right)^2 = -\frac{\delta N(\epsilon_F)}{N_0}\left(\frac{1}{1+KN(\epsilon_F)\tau_0}\right)^2 \quad (5)$$

In Eq.5, the coefficient is dependent on the density of state around the Fermi surface $N(\epsilon_F)$. In our experiment, the largest magnetoresistance at 15T is around 10% which means that the relative change of density of state $\delta N(\epsilon_F)/N(\epsilon_F)$ should be much smaller than 1 ($\frac{\delta N(\epsilon_F)}{N(\epsilon_F)} \ll 1$). Then, the Eq.(5) could approximately read as,

$$\begin{aligned}\frac{\delta\rho(B)}{\rho_0} &= -\frac{\delta N(\epsilon_F)}{N_0}\left(\frac{1}{1+KN(\epsilon_F)\tau_0}\right)^2 = -\frac{\delta N(\epsilon_F)}{N_0}\left(\frac{1}{1+KN_0\tau_0}\right)^2 \cdot \left(1 + \frac{\delta N(\epsilon_F)}{1+KN_0\tau_0}\right)^{-2} \\ &\approx -\frac{\delta N(\epsilon_F)}{N_0}\left(\frac{1}{1+KN_0\tau_0}\right)^2 \cdot \left(1 - \frac{2 \cdot \delta N(\epsilon_F)}{1+KN_0\tau_0}\right) \approx -\frac{\delta N(\epsilon_F)}{N_0}\left(\frac{1}{1+KN_0\tau_0}\right)^2 \quad (6)\end{aligned}$$

In the last equation, we neglect higher order terms and find the coefficient independent of external magnetic field. Our phenomenological model considers a special band structure illustrated in Fig. S4, which is a combination of parabolic bands and Dirac-type band. According to the detailed description in the main text, we notice the change of density of state in the Fermi surface would be linear to the external magnetic field, $\delta N(\epsilon_F) \propto B$. Therefore,

$$\frac{\delta\rho(B)}{\rho_0} \propto \frac{B}{N_0 \cdot (1+KN_0\tau_0)^2} \quad (7)$$

which clearly indicates the origin of the LMR.

## IV. The estimation of temperature-dependent mobility

Besides the qualitative descriptions mentioned in the main text, we present a quantitative estimation to show the consistency between data and the model. What we estimate is the temperature dependent mobility. As illustrated in Fig. S4, the energy dispersion could be expressed as

$$E = \begin{cases} -a\sqrt{\Delta_0^2 + k^2} + c, & E > E_0 \\ -bk^2 + d, & E < E_0 \end{cases}$$

where $a, b, c, d, \Delta_0$ are constants, $E_0$ is assumed as the boundary energy between these two bands. Since the density of state is $N(E) = \Omega \cdot k \left|\frac{dk}{dE}\right|$ where $\Omega$ is a constant, we can deduce that

$$N(E) = \begin{cases} \frac{\Omega}{a^2} \cdot E, & E > E_0 \\ \frac{\Omega}{2b}, & E < E_0 \end{cases}$$

As shown in the right side of Fig. S4, when the external magnetic field is provided, there would be a part of electrons occupying states in the linear band. We assume the proportion of such electrons is $\alpha$, so the total density of state at the Fermi level reads

$$N(E_F) = N_1(E) + N_2 = \alpha \cdot \frac{\Omega}{a^2} \cdot E + (1-\alpha)\frac{\Omega}{2b} \qquad (8)$$

The average mobility could be expressed by the average relaxation time

$$\mu_e = \frac{e\bar{\tau}}{m_e} \qquad (9)$$

The estimation of the average relaxation time $\bar{\tau}$ may be complex to formulize due to the complexity of the distribution function of electron. However, when we assume the Dirac distribution could be simply regarded as $\exp(-\frac{E}{k_B T})$, the average relaxation time could take a form equal to that of semiconductors [14]

$$\bar{\tau} \propto \frac{1}{k_B T} \frac{\int E\tau(E) \exp\left(-\frac{E}{k_B T}\right) N(E) dE}{\int \exp\left(-\frac{E}{k_B T}\right) N(E) dE} \qquad (10)$$

Eq.(10) could be applied to the electrons in our constructed special band structure since the Fermi level is assumed just around the top of the band which makes them behave like that in a semiconductor. However, the other parabolic bands, especially the large hole band around the $\Gamma$ point in lead films, could not be seen as such a simple case. But if we just consider samples in relatively low temperature, the difference between $1/(1 + \exp(E/k_B T))$ and $\exp(-E/k_B T)$ could be tolerable to make Eq.(10) applicable.

i)      For $N_1(E) = \alpha \cdot \frac{\Omega}{a^2} \cdot E$,

$$\bar{\tau}_{deformation1} \propto \frac{1}{k_B T} \frac{\int E \cdot \frac{a^2}{k_B T \alpha \Omega E} e^{-\frac{E}{k_B T}} \cdot \frac{\alpha \Omega E}{a^2} dE}{\int e^{-\frac{E}{k_B T}} \cdot \frac{\alpha \Omega E}{a^2} dE + \int e^{-\frac{E}{k_B T}} \cdot \frac{(1-\alpha)\Omega}{2b} dE}$$

$$= \frac{a^2}{k_B^2 T^2 \alpha \Omega} \cdot \frac{\int e^{-x} \cdot x\, dx}{\int e^{-x} \cdot x\, dx + d/T \int_0^\infty e^{-x} dx} = \frac{a^2}{k_B^2 \alpha \Omega (d * T + T^2)} \qquad (11)$$

where $x = \frac{E}{k_B T}$, $d = a^2(1-\alpha)/2\alpha b$.

ii)      For $N_2 = (1-\alpha)\frac{\Omega}{2b}$,

$$\bar{\tau}_{deformation2} \propto \frac{1}{k_B T} \frac{\int E \cdot \frac{2b}{k_B T(1-\alpha)\Omega} e^{-\frac{E}{k_B T}} \cdot \frac{(1-\alpha)\Omega}{2b} dE}{\int e^{-\frac{E}{k_B T}} \cdot \frac{\alpha \Omega E}{a^2} dE + \int e^{-\frac{E}{k_B T}} \cdot \frac{(1-\alpha)\Omega}{2b} dE}$$

$$= \frac{2b}{k_B T^2 (1-\alpha)\Omega} \frac{\int_0^\infty e^{-x} \cdot x\, dx}{\int e^{-x} \cdot x\, dx + d/T \int_0^\infty e^{-x} dx} = \frac{2b}{k_B (1-\alpha)\Omega(d*T + T^2)} \quad (12)$$

According to Eq.(11) and Eq.(12), the total mobility contributed by phonon-electron scattering has the relation:

$$\frac{1}{\bar{\tau}_{deformation}} \propto \frac{1}{\bar{\tau}_{deformation1}} + \frac{1}{\bar{\tau}_{deformation2}} = \frac{k_B^2 \alpha \Omega (d*T + T^2)}{a^2} + \frac{k_B (1-\alpha)\Omega(d*T + T^2)}{2b} = \frac{d*T + T^2}{M}$$

where $M$ is a constant. According to this function, the mobility could be then expressed as

$$\mu^{-1} = \mu_0^{-1} + \mu_{deformation}^{-1} = \mu_0^{-1} + C_1 T + C_2 T^2 \quad (13)$$

where $B$ and $C$ are constants, and $\mu_0$ refers to the contribution of other effects and is regarded as a constant. This is the approximately formulized function that can only be applicable when temperature is low enough and is usually not an exact estimation. Nevertheless, it is shown in Fig. S5 that our data could be well fitted below 50K in which LMR appears.

## V. The discussion on 20-ML Pb films grown on wetting layer and SIC phase interface

Figure S1(a) and S1(b) present the scanning tunneling microscopy (STM) images of 20-ML Pb films grown on SIC phase interface and amorphous wetting layer, respectively. As shown in Fig. S1(b), there are many line defects on the surface of 20-ML Pb film grown on wetting layer. Therefore, due to the amorphous interface between the film and the Si substrate, the Pb film on wetting layer exhibits a worse crystallinity, compared to the Pb film grown on crystalline SIC phase interface (Fig. S1(a)). This argument could be also justified by the larger residual resistance ratio (RRR) of Pb films on SIC phase interface (Fig S2(a) and S2(b)).

Figure S2(c)-(f) show the plot of $MR$ and $dMR/dB$ vs $B$ of 20-ML Pb films grown on amorphous wetting layer and SIC phase interface at various temperatures. The MR is increased by four times when the interfacial layer is changed from wetting layer to SIC phase. In Fig. S2(d), as magnetic field increases, the evolution of $dMR/dB$ of the films on SIC phase experiences three different regions marked as I, II and III. In this figure, the intercept of the $dMR/dB$ curve represents the linear terms ($MR \propto B$) while the gradient refers to the quadratic component ($MR \propto B^2$). Different regions represent different kinds of magnetoresistance, and thus a distinct transition from quadratic MR (region I with large slope and small intercept) to LMR (region III with small slope and large intercept) is revealed. However, we could not observe such clear transition in Fig. S2(c) since the curves do not show much difference at low fields and high fields especially when the temperature is above 20K. Therefore, it is distinguishable that the LMR is more predominant in Pb films grown on SIC phase.

It is noteworthy that both the 20-ML Pb films grown on SIC phase and wetting layer are protected by Si capping layer. However, the linear magnetoresistance only exists in 20-ML Pb film on SIC phase, indicating that the Si capping layer cannot lead to the observed linear magnetoresistance. Furthermore, previous works on Pb/Si systems reported that the Pb and Si do not form any stable silicides [16,17], which makes the Pb/Si system a feasible platform to study the physical properties of Pb films.

**VI. Figures**

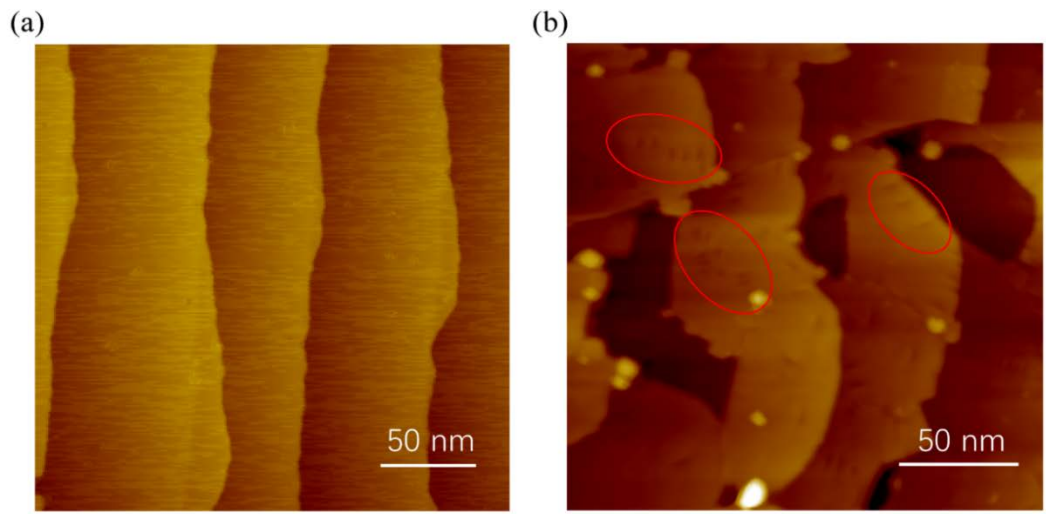

Fig. S1 (a) A typical STM image of 20-ML Pb film grown on SIC phase interface (250 nm × 250 nm). (b) A typical STM image of 20-ML Pb film grown on wetting layer (200 nm × 200 nm). The scale bars are 50 nm. There are many line defects mark as the red circles on the surface of 20-ML Pb film grown on wetting layer, indicating a worse crystallinity.

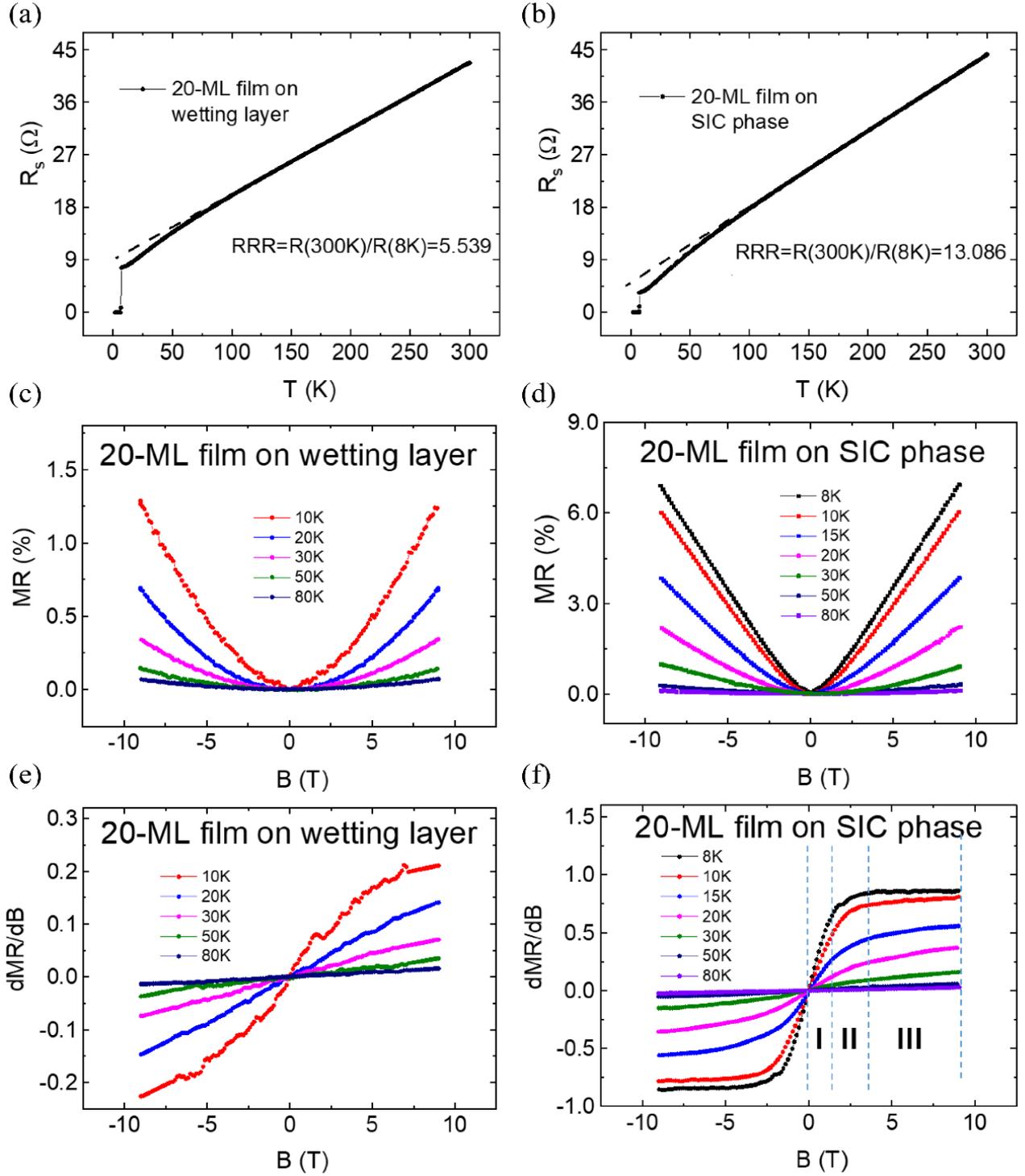

FIG. S2. Comparison between 20-ML Pb thin films grown on wetting layer and SIC phase interface. (a) (b) the temperature dependence of resistance from 2K to 300K. Residual resistance ratio (RRR) and magnetoresistance of Pb thin films grown on crystalline SIC phase are larger than that of Pb films on amorphous wetting layer, indicating that crystalline interface improves the quality of the Pb film. (c) (d) The overview of magnetoresistance ($MR = [R(H) - R(0)]/R(0)$) under perpendicular field for the films grown on (c) wetting layer and (d) SIC phase, respectively. (e)(f) The plot of $dMR/dB$ vs $B$ of 20-ML Pb films on (e) wetting layer and (f) SIC phase interface. The vertical blue dashed lines represent the borders of different regions marked as I, II and III.

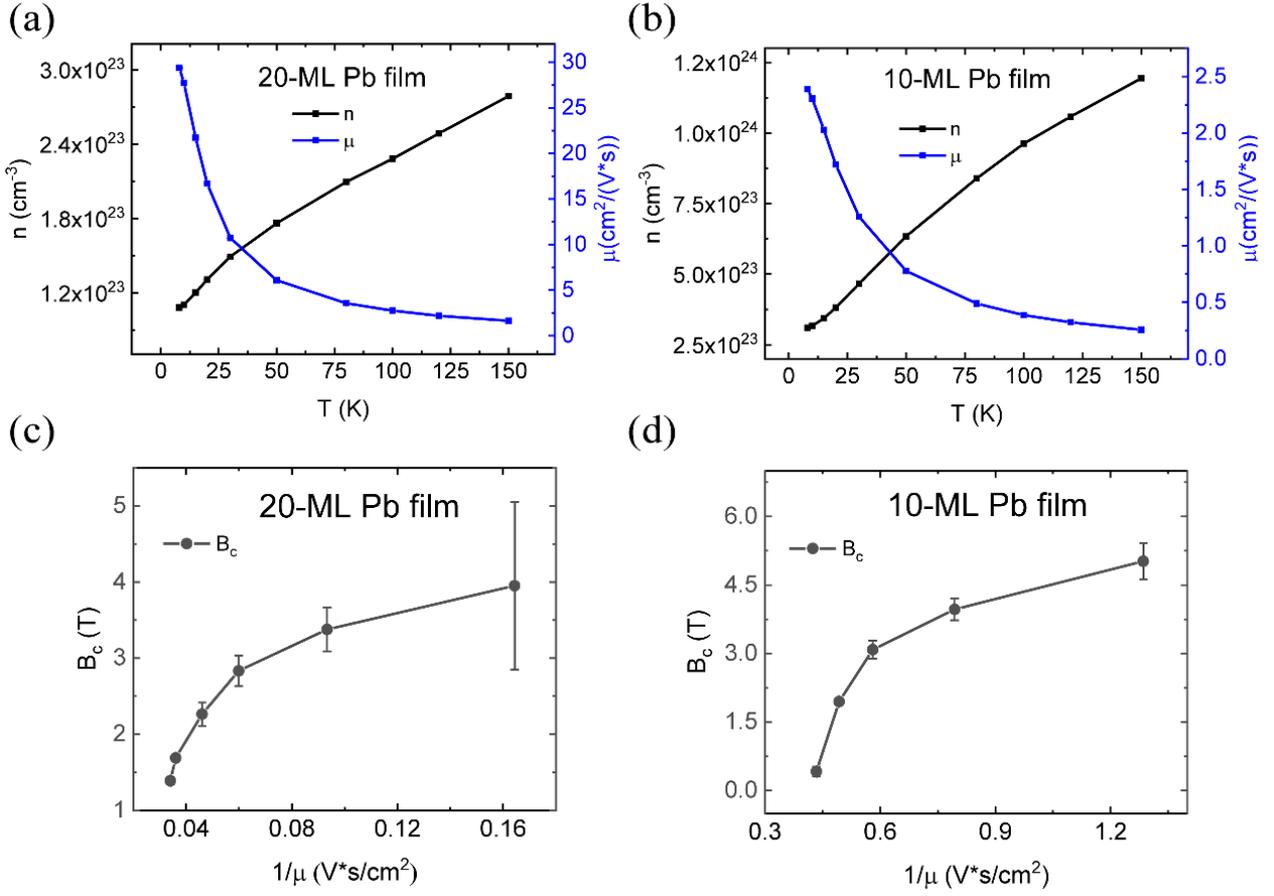

FIG. S3. The temperature dependent carrier density and mobility of (a) 20-ML and (b) 10-ML Pb thin films grown on SIC phase. The mobility of 20-ML Pb film is larger than that of 10-ML film due to lesser disorder. The curves of crossover field $B_c$ versus the inverse of mobility $1/\mu$ do not show a linear relationship for (c) 20-ML and (d) 10-ML Pb films inconsistent with the PL model.

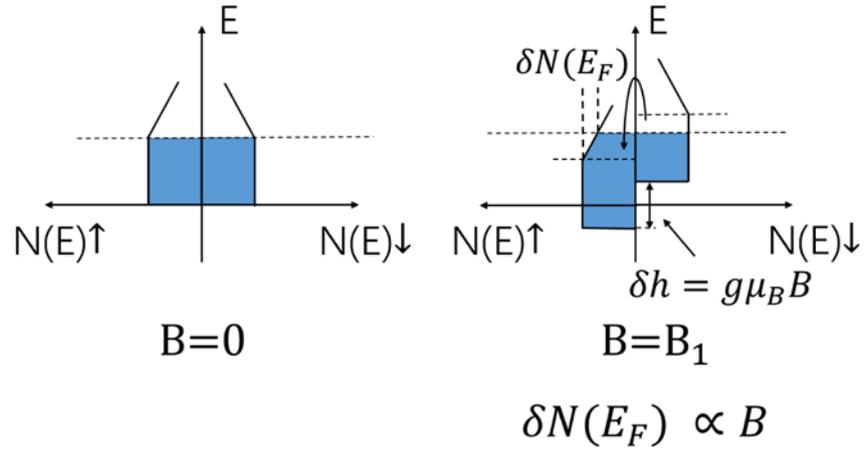

FIG. S4. The density of states (DOS) dependence of energy in the band structure consisting of linear (the oblique lines) and quadratic bands (the vertical lines) in two-dimensional system. The opposite arrows in the horizontal axis refer to opposite directions of the spins. The dashed line represents the Fermi level. The blue area represents the states occupied by carriers. As the magnetic field increases, the bands with different orientations of spins will shift oppositely, leading to the redistribution of carries indicated by the curved arrow.

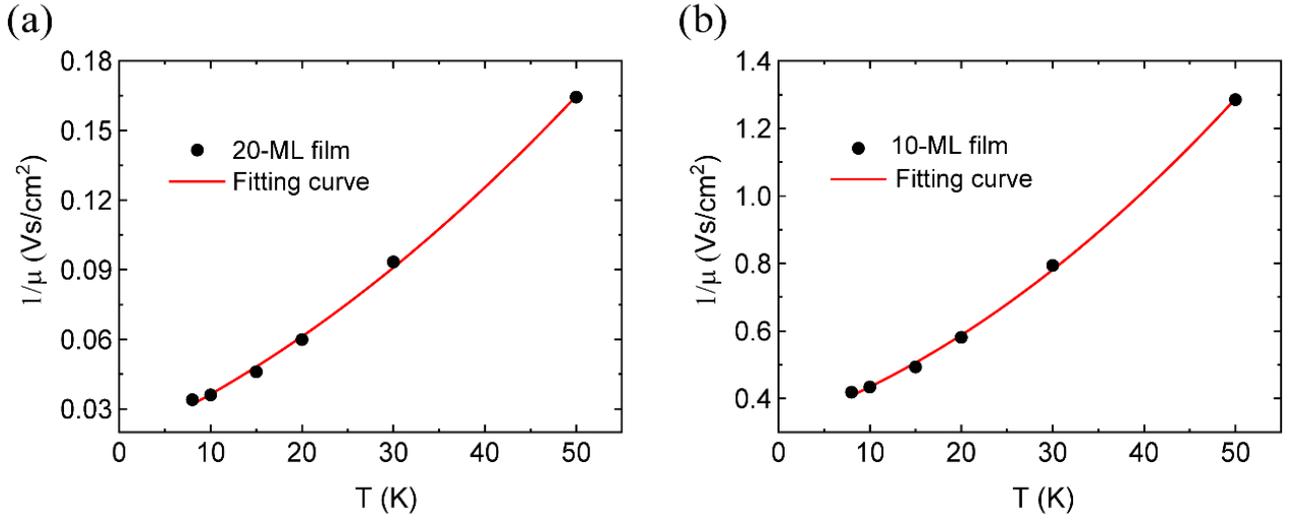

FIG. S5. The temperature dependent mobility of (a) 20-ML and (b) 10-ML Pb thin films on SIC phase, respectively. The red lines are the theoretical fitting curves (Eq.(13)). The fitting parameters $\mu_0^{-1} = (1.6 \pm 0.4) \times 10^{-2} (V \cdot s)/cm^2$, $C_1 = (1.76 \pm 0.33) \times 10^{-3} (V \cdot s)/(K \cdot cm^2)$, $C_2 = (2.40 \pm 0.06) \times 10^{-5} (V \cdot s)/(K^2 \cdot cm^2)$ for 20-ML film and $\mu_0^{-1} = (3.2 \pm 0.2) \times 10^{-1} (V \cdot s)/cm^2$, $C_1 = (9.3 \pm 1.8) \times 10^{-3} (V \cdot s)/(K \cdot cm^2)$, $C_2 = (2.0 \pm 0.3) \times 10^{-5} (V \cdot s)/(K^2 \cdot cm^2)$ for 10-ML film, respectively.

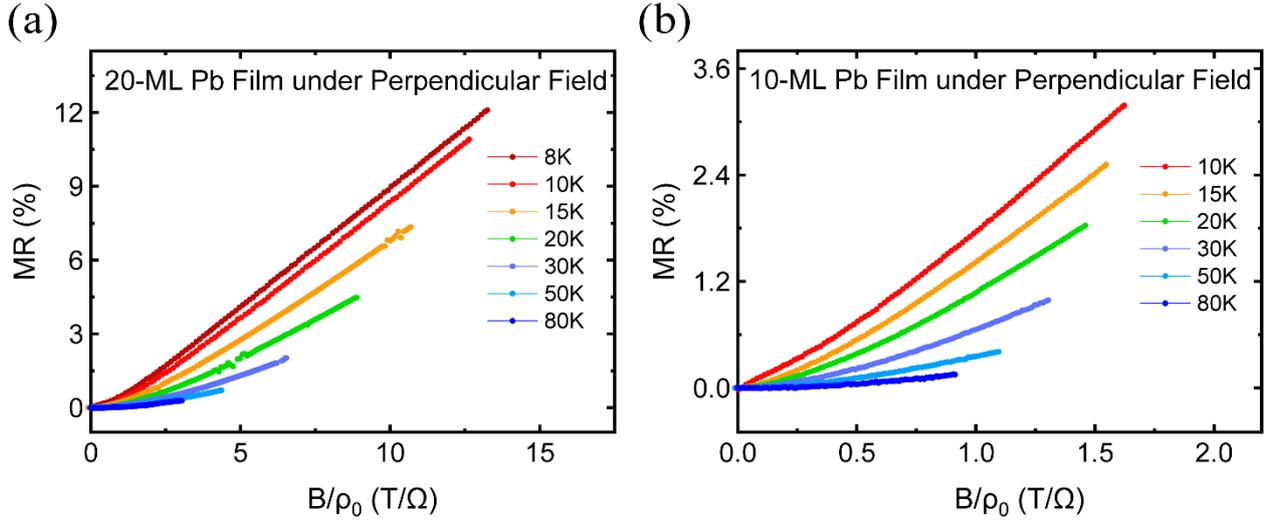

FIG. S6. The Kohler plots of (a) 20-ML Pb film and (b) 10-ML Pb film on SIC phase under perpendicular magnetic field.

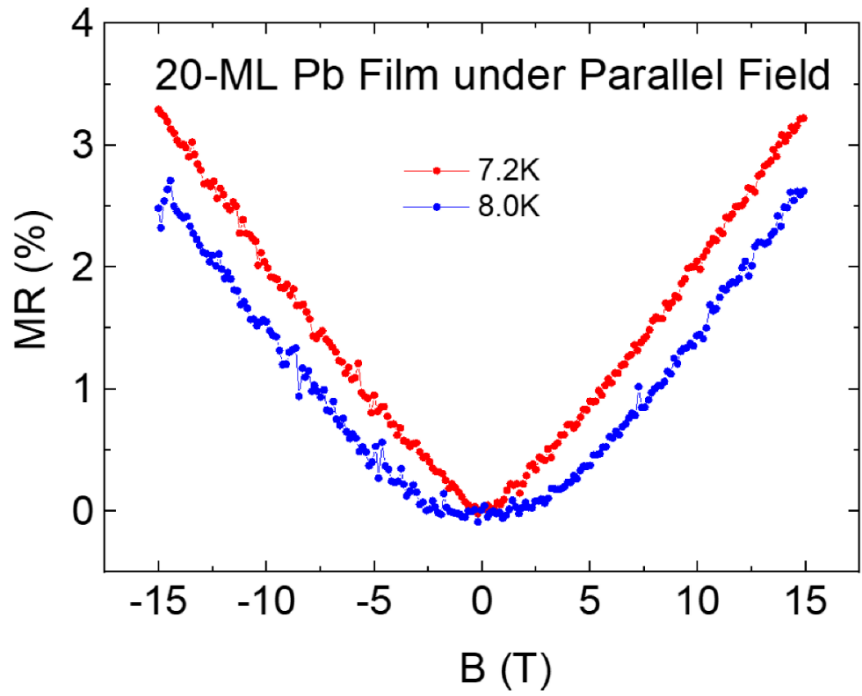

FIG. S7. The MR of 20-ML Pb film on SIC phase under parallel magnetic field.

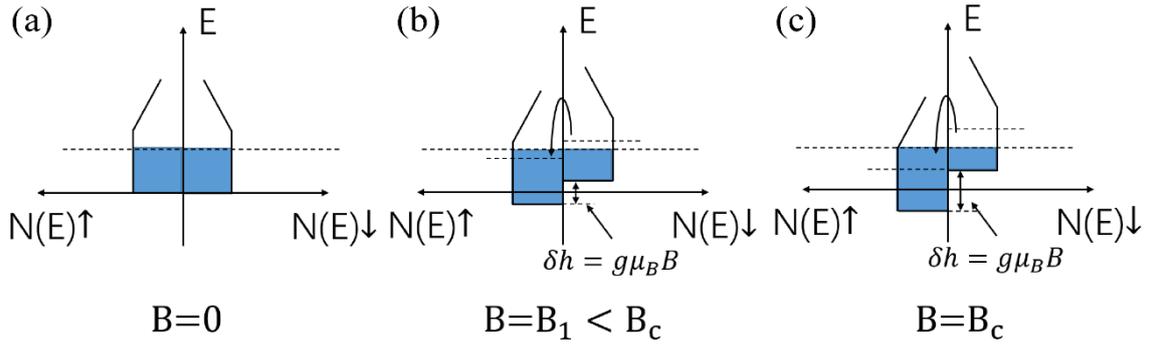

FIG. S8. The change from quadratic magnetoresistance to LMR. (a) the initial position of the Fermi surface, which is lower than the linear band area. (b) the DOS when the magnetic field B is lower than the crossover field $B_c$. All carriers are in the normal band area, so the Fermi surface remains unchanged. (c) the DOS when $B = B_c$. The Fermi surface reaches the linear band area, suggesting the appearance of linear magnetoresistance.

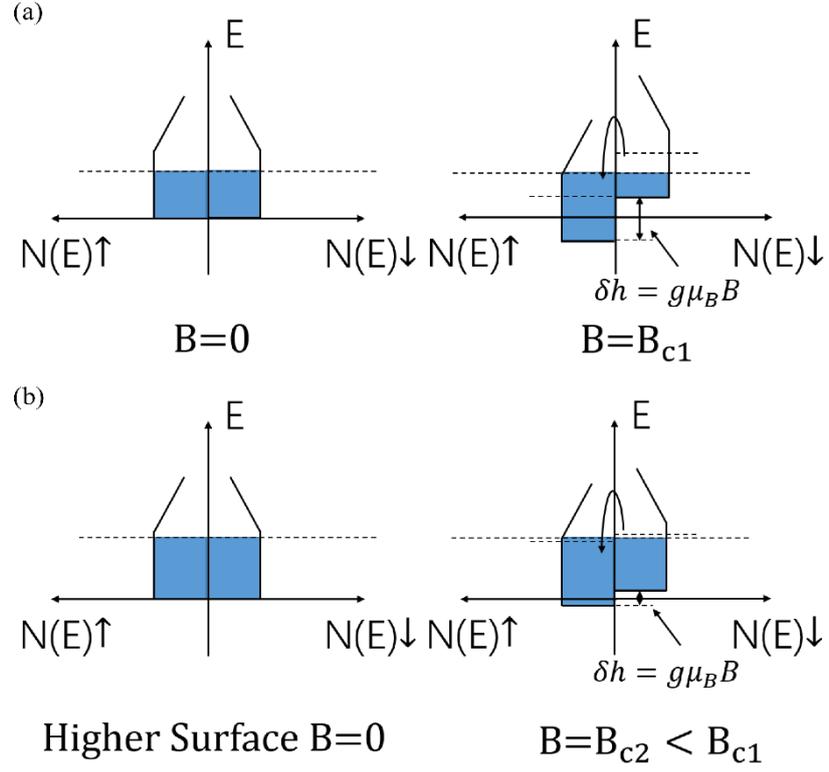

FIG. S9. The relation between crossover field $B_c$ and the initial position of the Fermi surface. The initial position of the Fermi surface of the bottom two pictures is lower than that of the upper two pictures. The shifts of the bands with opposite directions of spins clearly show the magnitude of the crossover field $B_c$. The lower Fermi surface requires a larger magnetic field to reach the region of linear bands.

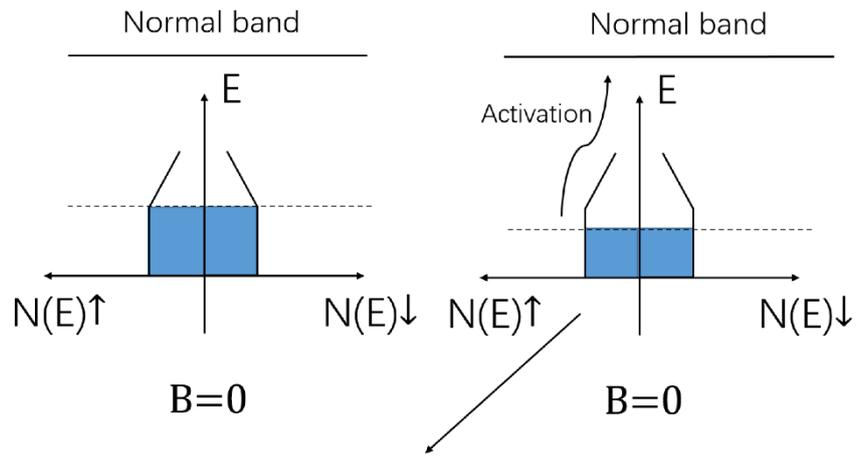

FIG. S10. Illustration of thermally activated behavior of the crossover field $B_c$. The initial position of the Fermi surface is shown in the left panel. Thermal activation brings carriers to the upper normal bands (the right panel). The dashed line becomes lower, indicating the loss of carriers, which requires a larger magnetic field to reach the linear bands and gives rise to LMR. Thus, $B_c$ increases with increasing temperature.

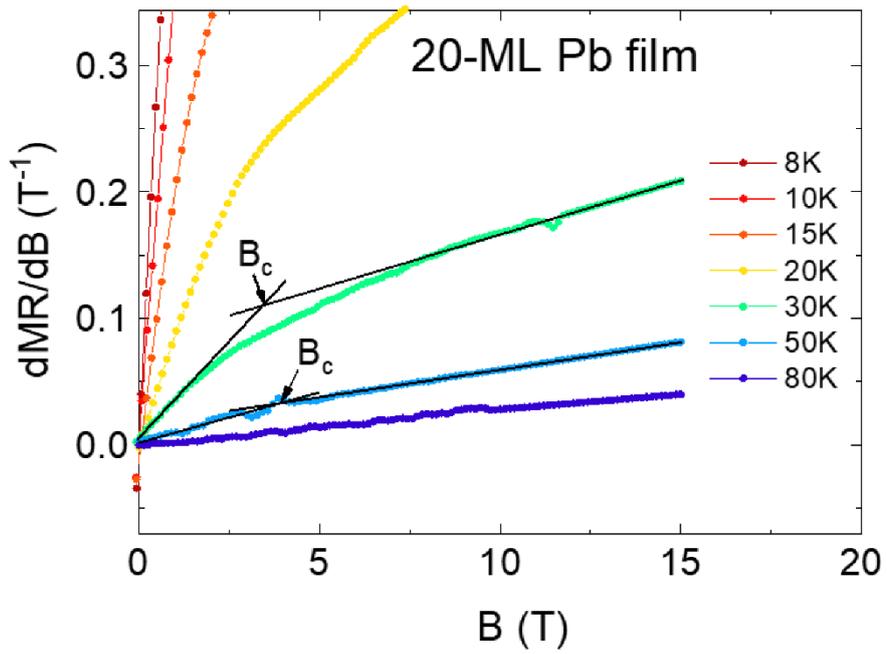

FIG. S11. The zoomed-in dMR/dB vs B plot of 20-ML Pb film on SIC phase. The crossover field is extracted from the crossing point of two linear fitting curves (black solid lines).

Y. Liu and Y. Tang contributed equally to this work.

*Corresponding author: Jian Wang (jianwangphysics@pku.edu.cn)

[1] A. Abrikosov, Phys. Rev. B **58**, 2788 (1998).

[2] A. Abrikosov, Europhys. Lett. **49**, 789 (2000).

[3] M. Parish and P. Littlewood, Nature **426**, 162 (2003).

[4] M. M. Parish and P. B. Littlewood, Nature **426**, 162 (2003).

[5] J. Hu and T. F. Rosenbaum, Nat. Mater. **7**, 697 (2008).

[6] Z. H. Wang, L. Yang, X. J. Li, X. T. Zhao, H. L. Wang, Z. D. Zhang, and X. P. A. Gao, Nano Lett. **14**, 6510 (2014).

[7] A. L. Friedman, J. L. Tedesco, P. M. Campbell, J. C. Culbertson, E. Aifer, F. K. Perkins, R. L. Myers-Ward, J. K. Hite, C. R. Eddy, G. G. Jernigan, and D. K. Gaskill, Nano Lett. **10**, 3962 (2010).

[8] W. Wang, Y. Du, G. Xu, X. Zhang, E. Liu, Z. Liu, Y. Shi, J. Chen, G. Wu, and X.-x. Zhang, Sci. Rep. **3**, 2181 (2013).

[9] J. M. Ziman, *Electrons and Phonons* (Oxford Univ. Press, 1960).

[10] D. Shoenberg, *Magnetic Oscillation in Metals* (Cambridge Univ. Press, 1984).

[11] A. A. Sinchenko, P. D. Grigoriev, P. Lejay, and P. Monceau, Phys. Rev. B **96**, 245129 (2017).

[12] Y. Feng, Y. Wang, D. M. Silevitch, J. Q. Yan, R. Kobayashi, M. Hedo, T. Nakama, Y. Ōnuki, A. V. Suslov, B. Mihaila, P. B. Littlewood, and T. F. Rosenbaum, PNAS **116**, 11201 (2019).

[13] B. Wu, V. Barrena, F. Mompeán, M. García-Hernández, H. Suderow, and I. Guillamón, Phys. Rev. B **101**, 205123 (2020).

[14] J. M. Ziman, *Principles of the Theory of Solids* (Cambridge. Univ. Press, 1972).

[15] J. Bardeen and W. Shockley, Phys. Rev. **80**, 72 (1950).

[16] A. E. Dolbak, R. A. Zhachuk, and B. Z. Olshanetsky, Cent. Eur. J. Phys. **2**, 254 (2004).

[17] R. P. Elliott, *Constitution of Binary Alloys,* (McGraw Hill, New York, 1965).